# Anti-scattering medium computational ghost imaging with modified Hadamard patterns

LI-XING LIN,[1] JIE CAO,[1,2,*] AND QUN HAO[1,3,4]

[1]*School of Optics and Photonics, Beijing Institute of Technology, Beijing 100081, China*
[2]*Yangtze Delta Region Academy, Beijing Institute of Technology, Jiaxing 314003, China*
[3]*Changchun University of Science and Technology, Changchun 130022, China*
[4]*qunhao@bit.edu.cn*
[*]*ajieanyyn@163.com*

**Abstract:** Illumination patterns of computational ghost imaging (CGI) systems suffer from reduced contrast when passing through a scattering medium, which causes the effective information in the reconstruction result to be drowned out by noise. A two-dimensional (2D) Gaussian filter performs linear smoothing operation on the whole image for image denoising. It can be combined with linear reconstruction algorithms of CGI to obtain the noise-reduced results directly, without post-processing. However, it results in blurred image edges while performing denoising and, in addition, a suitable standard deviation is difficult to choose in advance, especially in an unknown scattering environment. In this work, we subtly exploit the characteristics of CGI to solve these two problems very well. A kind of modified Hadamard pattern based on the 2D Gaussian filter and the differential operation features of Hadamard-based CGI is developed. We analyze and demonstrate that using Hadamard patterns for illumination but using our developed modified Hadamard patterns for reconstruction (MHCGI) can enhance the robustness of CGI against turbid scattering medium. Our method not only helps directly obtain noise-reduced results without blurred edges but also requires only an approximate standard deviation, i.e., it can be set in advance. The experimental results on transmitted and reflected targets demonstrate the feasibility of our method. Our method helps to promote the practical application of CGI in the scattering environment.

## 1. Introduction

Computational ghost imaging (CGI) is highly attractive with its flexible illumination design and single-pixel detection feature [1–3]. Different from the conventional thermal ghost imaging (GI) system [4,5], the modulation device used in CGI is programmable, such as a digital micromirror device (DMD) or spatial light modulator (SLM). Generally, the optical path in a CGI system can be divided into modulation device-to-object path (illumination optical path) and object-to-single-pixel detector path (detection optical path). Imaging through scattering medium techniques have a wide range of practical applications, and CGI has the nature of being immune to the scattering medium existing in the detection path. But the contrast of the illumination pattern reduces when passing through the scattering medium existing in the illumination optical path, which causes the effective information in the reconstruction result to be drowned out by noise [6–9].

Hadamard-based CGI (called HCGI in this paper) [10] performs better in noisy and scattering environments compared to random and Fourier basis patterns because of its differential measurement feature, as well as the orthogonality and optimizable ordering of Hadamard basis patterns [11–13]. Recently, some Hadamard-based anti-scattering CGI methods have been proposed to further improve its anti-scattering ability [14–16], but they rely on the help of additional measurement techniques, such as optical polarization [14], measuring transmission matrix [15], and deep learning algorithms [16].

One of the advantages of CGI is that patterns used for illumination and reconstruction can be set differently to directly achieve specific imaging purposes, without post-processing, such as super-resolution [17], edge detection, and anti-noise [18,19]. The 2D Gaussian filtering [20] is suitable for direct combination with linear reconstruction algorithms of CGI to obtain noise-reduced results since it performs a linear operation on the image [18,19]. However, it results in blurred image edges while performing denoising and, in addition, it is difficult but necessary to choose a suitable standard deviation in advance in an unknown scattering environment. In this work, we demonstrate that these two problems can be solved very well by exploiting the designable mapping relationship between patterns used for illumination and image reconstruction of CGI and combining them with the differential operation features of HCGI.

Specifically, we develop modified positive and negative Hadamard patterns based on a 2D Gaussian filter and obtain a corresponding modified Hadamard sampling matrix after the differential operation. Using the modified Hadamard sampling matrix for image reconstruction allows HCGI perform well against the influence of turbid scattering medium. The method of generating modified Hadamard patterns is described in section 2. We analyze and demonstrate that using Hadamard patterns for illumination but using our developed modified Hadamard patterns for reconstruction (called MHCGI in this paper) can enhance the robustness of CGI against turbid scattering medium. Our method not only helps directly obtain noise-reduced results without blurred edges but also requires only an approximate standard deviation, which can be set in advance. Experimental results show that the image quality is considerably improved.

## 2. Method

### 2.1 Generation approach of our modified Hadamard patterns

Our modified Hadamard patterns are generated in the following three steps. Taking one of the series patterns as an example, the flow chart is shown in Fig. 1.

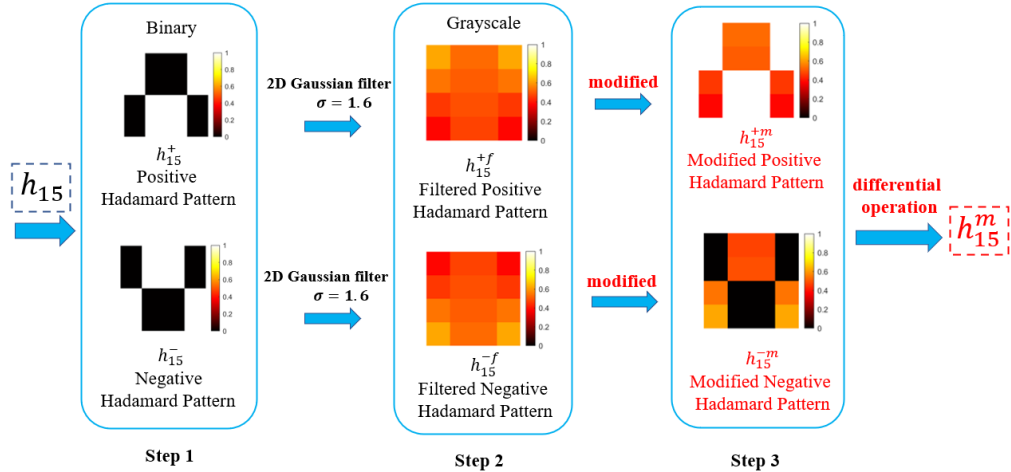

Fig. 1. The flow chart for generating one of the modified Hadamard patterns used for image reconstruction. $h_{15}$ is a $4 \times 4$ Hadamard pattern containing elements value of $\pm 1$, $h_{15}^m$ is the generated modified Hadamard pattern, which is also a part of the corresponding modified Hadamard sampling matrix. Step 3 is the most important part of our approach (marked red).

Step 1: generate a series of Hadamard patterns to be loaded into the DMD for illumination as follows:
The Natural order Hadamard matrix can be obtained by the following function:

$$H_{\mathrm{N}} = H_{2^n} = H_2 \otimes H_{2^{n-1}} = \begin{bmatrix} H_{2^{n-1}} & H_{2^{n-1}} \\ H_{2^{n-1}} & -H_{2^{n-1}} \end{bmatrix} \qquad (1)$$

where $H_1 = 1$, $2^n$ is the order of the Hadamard matrix, $n$ is a positive integer (i.e., $2^1$ is the lowest order), and $\otimes$ is the Kronecker product. Assuming that, $N = K \times K$, then each row of the Hadamard matrix $H_N$ can be reshaped into a $K \times K$ matrix as a Hadamard pattern containing only values of $\pm 1$. Taking $N = 2^n = 16$ (K=4) as an example:

$$H_4 = \begin{bmatrix} 1 & 1 & 1 & 1 \\ 1 & -1 & 1 & -1 \\ 1 & 1 & -1 & -1 \\ 1 & -1 & -1 & 1 \end{bmatrix} \tag{2}$$

each row of $H_4$ can be reshaped into a $2 \times 2$ matrix:

$$h_1 = \begin{bmatrix} 1 & 1 \\ 1 & 1 \end{bmatrix}, h_2 = \begin{bmatrix} 1 & 1 \\ -1 & -1 \end{bmatrix}, h_3 = \begin{bmatrix} 1 & -1 \\ 1 & -1 \end{bmatrix}, h_4 = \begin{bmatrix} 1 & -1 \\ -1 & 1 \end{bmatrix} \tag{3}$$

however, DMD is a binary ({0,1}) modulator, to realize +1 and -1 modulation, every single Hadamard pattern ($h_i$) needs to be divided into a pair of binary patterns (0 or 1), $h_i^+$ and $h_i^-$, which are illumination patterns to be loaded into DMD, then $h_i$ can be obtained by performing the differential operation:

$$h_i = h_i^+ - h_i^- \tag{4}$$

where $h_i^+ = (h_i + 1_i)/2$, $h_i^- = (1_i - h_i)/2$, and $1_i$ is a $K \times K$ matrix of ones. Taking $h_2$ as an example:

$$h_2^+ = \begin{bmatrix} 1 & 1 \\ 0 & 0 \end{bmatrix}, h_2^- = \begin{bmatrix} 0 & 0 \\ 1 & 1 \end{bmatrix}, h_2 = \begin{bmatrix} 1 & 1 \\ 0 & 0 \end{bmatrix} - \begin{bmatrix} 0 & 0 \\ 1 & 1 \end{bmatrix} \tag{5}$$

that is why differential measurements in Hadamard-based CGI are needed.

Step 2, generate filtered Hadamard patterns as follows:
Performing 2D Gaussian filter with standard deviation $\sigma$ on $h_i^+$ and $h_i^-$, then a series of filtered patterns which are grayscale are obtained. Denoting filtered $h_i^+$ as $h_i^{+f}$, $h_i^-$ as $h_i^{-f}$, and $h_i$ as $h_i^f$. $h_i^f = h_i^{+f} - h_i^{-f}$, and denoting the corresponding sampling matrix as $FH_N$. To better illustrate, taking the reshaped 15$^{th}$ row of $H_{16}$ (K=4), $h_{15}$ (a $4 \times 4$ matrix), as an example here:

$$h_{15} = \begin{bmatrix} 1 & -1 & -1 & 1 \\ 1 & -1 & -1 & 1 \\ -1 & 1 & 1 & -1 \\ -1 & 1 & 1 & -1 \end{bmatrix}, h_{15}^+ = \begin{bmatrix} 1 & 0 & 0 & 1 \\ 1 & 0 & 0 & 1 \\ 0 & 1 & 1 & 0 \\ 0 & 1 & 1 & 0 \end{bmatrix}, h_{15}^- = \begin{bmatrix} 0 & 1 & 1 & 0 \\ 0 & 1 & 1 & 0 \\ 1 & 0 & 0 & 1 \\ 1 & 0 & 0 & 1 \end{bmatrix} \tag{6}$$

when the standard deviation of the filter is 1.6, we have a pair of filtered patterns:

$$h_{15}^{+f} = \begin{bmatrix} 0.6188 & 0.5289 & 0.5289 & 0.6188 \\ 0.5449 & 0.5109 & 0.5109 & 0.5449 \\ 0.4551 & 0.4891 & 0.4891 & 0.4551 \\ 0.3812 & 0.4711 & 0.4711 & 0.3812 \end{bmatrix}, h_{15}^{-f} = \begin{bmatrix} 0.3812 & 0.4711 & 0.4711 & 0.3812 \\ 0.4551 & 0.4891 & 0.4891 & 0.4551 \\ 0.5449 & 0.5109 & 0.5109 & 0.5449 \\ 0.6188 & 0.5289 & 0.5289 & 0.6188 \end{bmatrix} \quad (7)$$

and $h_{15}^f$:

$$h_{15}^f = \begin{bmatrix} 0.2376 & 0.0578 & 0.0578 & 0.2376 \\ 0.0898 & 0.0218 & 0.0218 & 0.0898 \\ -0.0898 & -0.0218 & -0.0218 & -0.0898 \\ -0.2376 & -0.0578 & -0.0578 & -0.2376 \end{bmatrix} \quad (8)$$

Step 3, generate our modified Hadamard patterns used for image reconstruction. It is important to emphasize that this step is the core of our proposed approach.

Setting the value of the element in the matrix $h_i^{+f}$ corresponding to the position of element +1 in the matrix $h_i^+$ to +1, the modified positive Hadamard patterns are obtained, denoted as $h_i^{+m}$. And setting the value of the element in the matrix $h_i^{-f}$ corresponding to the position of element 0 in the matrix $h_i^-$ to 0, the modified negative Hadamard patterns are obtained, denoted as $h_i^{-m}$. Then our modified Hadamard patterns are obtained by performing the differential operation on $h_i^{+m}$ and $h_i^{-m}$, that is, $h_i^m = h_i^{+m} - h_i^{-m}$, and denoting the corresponding modified Hadamard matrix as $MH_N$, which is only used for image reconstruction in MHCGI (our method). Note that the "Cake-cutting" ordering [21] of the Hadamard matrix is used in our experiments and reconstruction algorithm, and we explain this in the next section.

$$h_{15}^{+m} = \begin{bmatrix} 1 & 0.5289 & 0.5289 & 1 \\ 1 & 0.5109 & 0.5109 & 1 \\ 0.4551 & 1 & 1 & 0.4551 \\ 0.3812 & 1 & 1 & 0.3812 \end{bmatrix}, h_{15}^{-m} = \begin{bmatrix} 0 & 0.4711 & 0.4711 & 0 \\ 0 & 0.4891 & 0.4891 & 0 \\ 0.5449 & 0 & 0 & 0.5449 \\ 0.6188 & 0 & 0 & 0.6188 \end{bmatrix} \quad (9)$$

and $h_{15}^m = h_i^{+m} - h_i^{-m}$:

$$h_{15}^m = \begin{bmatrix} 1 & 0.0578 & 0.0578 & 1 \\ 1 & 0.0218 & 0.0218 & 1 \\ -0.0898 & 1 & 1 & -0.0898 \\ -0.2376 & 1 & 1 & -0.2376 \end{bmatrix} \quad (10)$$

The 2D Gaussian filter smooths the whole image, which may cause the high-energy components of the pattern to be attenuated as well. However, by our modified operation, elements 1 in $h_i^{+m}$ and elements 0 in $h_i^{-m}$ ensure the fluctuation properties while filtering. Those elements with a value of 1 in $h_i^m$ make it achieves filtering while retaining the original high-energy components in the sampling matrix (comparing Eq. (8) with Eq. (10) as an example). This helps MHCGI reduce noise without blurring image edges. Also owing to this modified operation and the differential operation, the filtered components retained in the sampling matrix do not differ much for larger standard deviations. This makes MHCGI less stringent in selecting the standard deviation. We prove this feature in Section 3, as shown in Fig. 3 and Fig. 5.

## 2.2 Image Reconstruction principle of MHCGI

Nature ordering of Hadamard basis patterns allows for a theoretically perfect reconstruction of the image at the full sampling ratio. But it takes a long time. Different sensing basis ordering methods of the Hadamard matrix have been proposed to reconstruct the high-quality image at low sampling rates, such as "Cake-Cutting" ordering (CC) [21], Walsh ordering [22], and "Russian Doll" ordering [23]. We choose the CC ordering of the Hadamard matrix in our experiments and reconstruction algorithm, as it performs well against noise and can be easily combined with the structured character of the Hadamard matrix to accelerate the computational process [15,24,25].

An HCGI system acquires target information by projecting a series of ordered Hadamard patterns onto an object and simultaneously detecting the corresponding reflected or transmitted total light intensity values with a single-pixel detector. Here, the intensity distribution of the object is expressed as a $n \times n$ matrix $T$, the $i^{th}$ projected pattern is expressed as an $n \times n$ matrix $h_i^+$ and $h_i^-$ (the same meaning as that in Eq. (4)), and the corresponding detected value is expressed as $B_i^+$ and $B_i^-$, and $B_i = B_i^+ - B_i^-$ (the differential measurement in HCGI).

Permutating the matrix $T$ into an N×1 column vector $X$ (N $= n \times n$). After 2N detection from the single-pixel detector, and performing the differential measurement, all bucket values can be permutated into an N×1 column vector $Y = [B_1 \cdots B_s \cdots B_N]^T$. Then the imaging principle of HCGI at full sampling ratio can be modeled as a linear process as follows:

$$Y = H_N X \tag{11}$$

where $H_N$ is the Nature order Hadamard matrix, called sampling matrix in the imaging model, which has the same meaning as that in Eq. (1). Since $H_N$ is an orthogonal matrix, $(H_N)^T H_N = qI$ holds, where $(H_N)^T$ denotes the transposition of the matrix $H_N$, $I$ is the identity matrix of size $N \times N$, and $q$ is a real constant equal to $N$. $X$ is obtained:

$$X = \frac{1}{N}(H_N)^T Y \tag{12}$$

Considering the low sampling ratio situation, $H_N$ is ordered according to the CC ordering method, denoted as $H_{NC}$. That is, illumination patterns used in our experiments are CC ordering of Hadamard patterns. Assuming that the sampling number is $K$ which is less than $N$, we obtain $Y_{NCK} = [B_1 \cdots B_K \cdots B_N]^T$ ($K < N$), and $[B_{K+1} \cdots B_N]^T = 0$. To still solve for $X$ using the imaging model of Eq. (12), let the elements of rows $K$ through $N$ of the sampling matrix $H_{NC}$ be equal to zero, denoted as $H_{NCK}$. We have:

$$Y_{NCK} = H_{NCK} X \rightarrow X = \frac{1}{N}(H_{NCK})^T Y_{NCK} \tag{13}$$

called HCGI in this paper.

If the 2D Gaussian filtered (with standard deviation, σ) Hadamard patterns (denoting the corresponding sampling matrix as $FH_{NCK}$) are used for image reconstruction (called FHCGI in this paper), the reconstruction result is:

$$X_F = \frac{1}{N}(FH_{NCK})^T Y_{NCK} \tag{14}$$

by shaping the matrix $X_F$ back to an $n \times n$ matrix $T_F$, a filtered (noise-reduce) image is reconstructed directly. But we experimentally find the image edges are blurred, and the larger the stand deviation, the more pronounced the blurred edges. Besides, it is difficult to choose a suitable value of σ in advance in an unknown scattering environment.

In our proposed MHCGI scheme, the CC ordering of Hadamard patterns (the corresponding sampling matrix is $H_{NCK}$) are used for illuminating the object and the modified Hadamard patterns (the corresponding sampling matrix is $MH_{NCK}$) are used for image reconstruction, we have the image reconstruction algorithm of MHCGI:

$$X_M = \frac{1}{N}(MH_{NCK})^T Y_{NCK} \qquad (15)$$

by reshaping the matrix $X_M$ back to an $n \times n$ matrix $T_M$, a noise-reduce without blurred edges image can be reconstructed directly when existing turbid scattering medium in the illumination optical path. And a slightly larger value of σ ensures that the MHCGI works under different scattering intensity conditions, that is, the stand deviation can be easily chosen in advance. This advantage is attributed to the modified method described in Step 3 of Section 2.1.

## 3. Experiments and results

The schematic diagram of the experimental setup is shown in Fig. 2. A beam of white light (emitted by an optical fiber white LED light source) beams onto the DMD to sequentially generate a series of modulated designed patterns. Every illumination pattern passes through the scattering medium and illuminates the reflective object. The corresponding reflective light passes through the scattering medium and detected by a single-pixel detector (SPD). We use different concentrations' emulsions to simulate the effect of scattering medium with different scattering intensities. The emulsion is created by mixing a certain amount of whole milk (GB25190) with pure water. A transparent container containing 256.5 (2.5cm×11.4cm×9.0cm) millimeters of purified water is placed in the illumination optical path. 0, 0.2, 0.4, 0.6, 0.8, and 1.0 millimeters (ml) of whole milk are respectively added to the container to represent different scattering intensities of the scattering medium. Ambient light is kept on during the whole experiment to introduce ambient noise. For more visualization, we record some scenes of a Hadamard pattern passing through a turbid scattering medium with different scattering intensities, as shown in Fig. S1 and Fig. S2 in Supplement 1. (See Supplement 1 for details). The gain of the single-pixel detector and the power of the LED are both constant in the experiments with different scattering intensities.

The image size of every pattern is 64×64 pixels. The pixel size of the DMD is 13.68um×13.68um, and 12×12 pixels (164.16 um×164.16 um) are combined to produce one pixel of the designed pattern. A combined projection lens L1 with a magnification of 5.71× is used to project generated patterns onto a reflective object.

The root mean square error (RMSE) (defined as Eq. (16)), peak signal-to-noise ratio (PSNR) (defined as Eq. (17)), and the structural similarity (SSIM) index [26] of the reconstruction result is calculated to quantitatively analyze the image quality.

$$RMSE = \left[\frac{1}{nm}\sum_{i,j=1}^{m,n}(I_1(i,j)-I_0(i,j))^2\right]^{1/2} \qquad (16)$$

$$PSNR = 10\log_{10}\left(peakval^2/RMSE^2\right) \qquad (17)$$

where $I_0$ and $I_1$ are 2D images of size $m \times n$ pixels; '*peakval*' is the maximum pixel value. The RMSE measures the difference between a reference image $I_0$ and a restored image $I_1$. The lower the RESE value, the better the quality of the restored image. And the higher the PSNR, the better the quality of the restored image. The unit of PSNR is dB.

The SSIM($x$, $y$) is calculated as follows:

$$SSIM(x, y) = \frac{(2\mu_x\mu_y + c_1)(2\sigma_{xy} + c_2)}{(\mu_x^2 + \mu_y^2 + c_1)(\sigma_x^2 + \sigma_y^2 + c_2)} \tag{18}$$

where $x$ stands for the restored image and $y$ stands for the reference image; $\mu_x$ ($\mu_y$) and $\sigma_x^2$ ($\sigma_y^2$) denote the mean and variance of $x$ ($y$), respectively; $\sigma_{xy}$ is the covariance of $x$ and $y$; and $c_1$ and $c_2$ are constants to stabilize the division with a weak denominator. The value of SSIM ranges from 0 to 1, and the higher the SSIM value, the better the similarity between the restored image and the reference image.

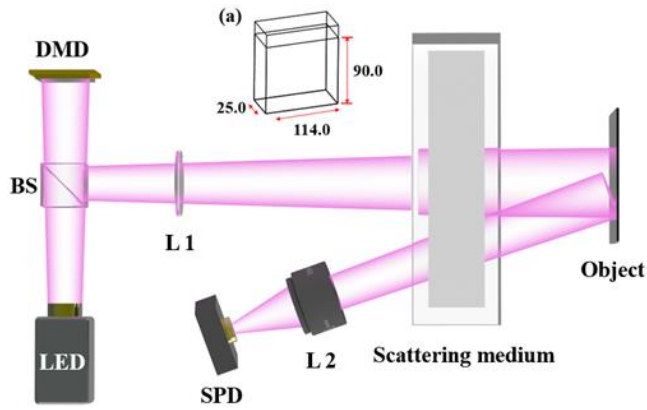

Fig. 2. Schematic diagram of the experimental setup. BS: beam splitter; L1: a combined projection lens with magnification 5.71×; L2: a focusing lens; Object: reflective target; The object is affixed to the back of the container. SPD: single-pixel detector; (a): physical size details of the transparent container for scattering medium. The unit of length is millimeter (mm).

Experimental results of using different sampling matrices for image reconstruction at the sampling ratio of 0.75 are shown in Figs. 3-6. The image size is 64×64 pixels. All reconstructed images are normalized to 0 to 1. $H_{NCK}$, $FH_{NCK}$, and $MH_{NCK}$ are the sampling matrices used for the image reconstruction algorithm corresponding to Eq. (13), Eq. (14), and Eq. (15), respectively. Three types of reflective objects with different levels of complexity are used: Chinese characters, a negative resolution target (3in × 3in), and the grayscale 'Lena' image. Note that because of the limited imaging field of view, a portion of the resolution target was imaged in the experiment. Ambient light is kept on during the experiment to introduce ambient noise.

### 3.1 Experimental results of FHCGI and MHCGI with different standard deviations of the 2D Gaussian filter

The standard deviation is important for the denoising effect of the 2D Gaussian filter, so the natural question is whether our method is very strict about the selection of the standard deviation. We demonstrate the reconstruction results of FHCGI and MHCGI with different standard deviations of the 2D Gaussian filter at no scattering condition, 0.0ml of whole milk is added to the container, as shown in Fig. 3.

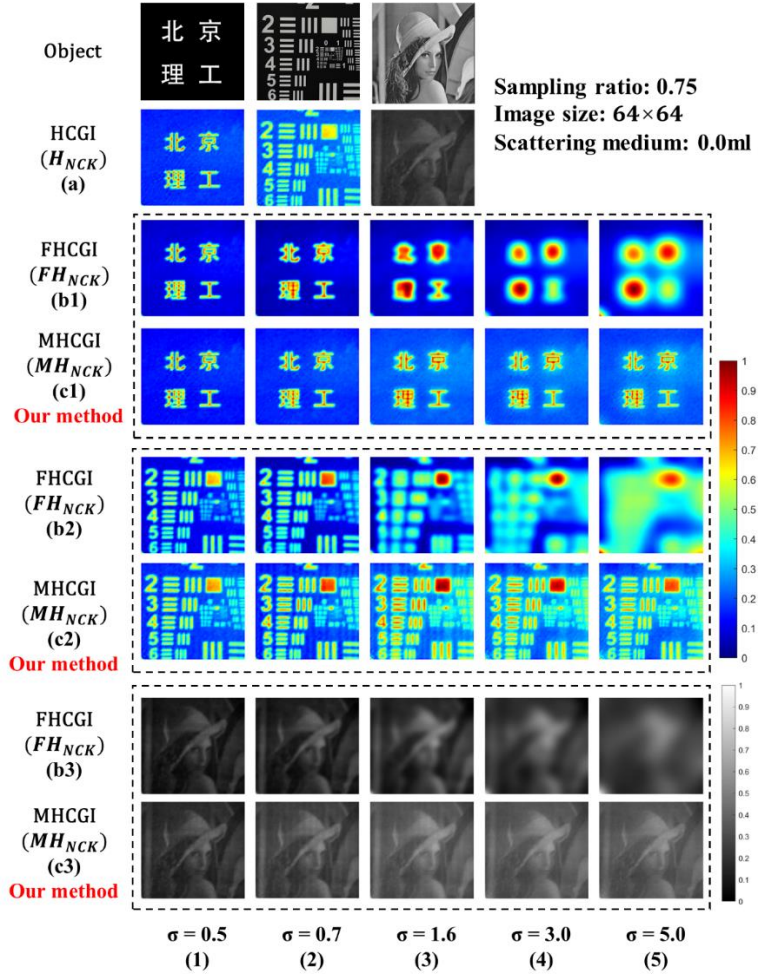

Fig. 3. Reconstruction results of FHCGI and MHCGI with different standard deviations of the 2D Gaussian filter at no scattering condition. The sampling ratio is 0.75. Ambient light is kept on during the experiment to introduce ambient noise. Object: three types of reflective objects with different levels of complexity. (a): the reconstruction results of HCGI. (b1), (b2) and (b3): the reconstruction results of FHCGI. (c1), (c2), and (c3): the reconstruction results of MHCGI (our method, marked in red). (1)-(5): the reconstruction results when the standard deviation of the 2D Gaussian filter is 0.5, 0.7, 1.6, 3.0, and 5.0, respectively.

According to Fig. 3(b1), (b2), and (b3), obviously, as the standard deviation increases, the reconstructed image quality of FHCGI becomes gradually better and then gradually worse, as it leads to severe edge blurring when the standard deviation is larger, which means that FHCGI is more sensitive to changes in standard deviation. That is, it is more stringent in selecting standard deviation. In contrast, the image reconstructed by MHCGI does not change much with increasing standard deviation. That is, MHCGI is insensitive to changes in standard deviation.

## 3.2 Experimental results of HCGI, FHCGI, and MHCGI under different scattering intensities

The reconstruction results under different scattering intensities at the sampling ratio of 0.75 are shown in Fig. 4. According to Fig. 3(b1), (b2), and (b3), FHCGI performs well when the standard deviation is 0.7. Therefore, we choose the imaging results when the standard deviation of the 2D Gaussian filter of both FHCGI and MHCGI is 0.7 as a comparison. The object is a negative resolution target and the imaging field of view is part of it, as shown in the first row of Fig. 4. 0, 0.2, 0.4, 0.6, 0.8 and 1.0 millimeters (ml) of whole milk are respectively added to the container to represent different scattering intensities of the scattering medium. Ambient light is kept on during the experiments. SSIM and PSNR marked below each image are used as the evaluation metrics.

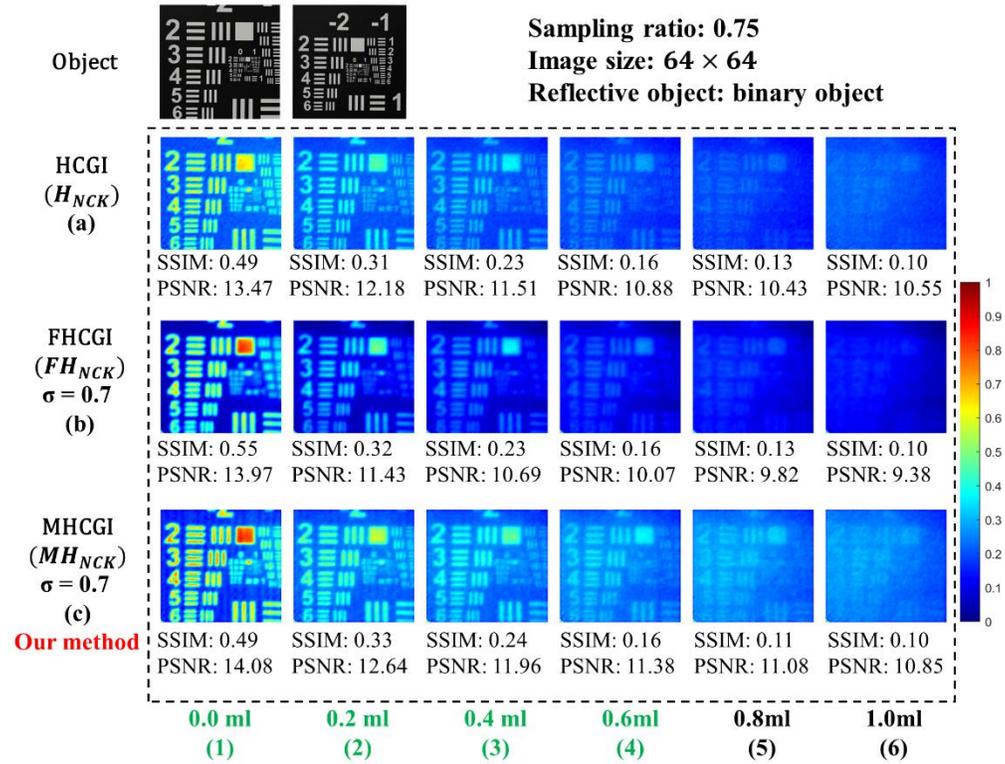

Fig. 4. Reconstruction results at the sampling ratio of 0.75. Ambient light is kept on during the experiments. (a), (b), and (c): reconstruction results of HCGI, FHCGI, and MHCGI, respectively. The standard deviation of the 2D Gaussian filter is 0.7. The image size is 64×64 pixels. (1)-(6): 0.0ml (only pure water in the container), 0.2ml, 0.4ml, 0.6ml, 0.8, and 1.0ml of whole milk (GB25190) are respectively added to the container in front of the object. The value of SSIM and PSNR are marked below each image.

In general, the imaging quality of all three methods gradually decreases as the scattering intensity increases. Three methods work under no-scattering and moderate-scattering conditions, as shown in Fig. 4(1)-(4), marked green. But by strong scattering, all barely reconstruct, as shown in Fig. 4(5) and 4(6). Comparing Fig. 4(b) and 4(a), although FHCGI has a better effect of reducing background noise than HCGI, FHCGI causes blurring of image edges if the standard variance is not chosen properly. But our proposed MHCGI performs best in reducing background noise without causing edge blurring, comparing Fig. 4(c) with Fig. 4(b) and Fig. 4(a).

Fig. 5(a) shows curves of PSNR of the MHCGI reconstructed images at different standard deviations (0.5, 0.7, 1.6, 3, 5, 8, and 10) under different scattering intensities. Fig. 5(b) shows curves of PSNR of HCGI and MHCGI at different scattering intensities. The object is the negative resolution target. As the standard deviation increases, the enhancement of the useful signal by MHCGI increases significantly and then flattens out (as shown in Fig. 5(a)). The reason is that the sampling matrix $MH_{NCK}$ remains essentially constant when the standard deviation is greater than 5. Besides, in no scattering condition (low noise level environment), HCGI performs well, but MHCGI with large standard deviation (e.g. $\sigma = 8, 10$) will instead lead to a slight degradation of the image quality (as shown in Fig. 5(b)).

In short, MHCGI performs better than HCGI in a scattering environment. The advantage of MHCGI is that it allows for a wider choice of standard deviations. A slightly larger value of standard deviation (such as 0.7-5.0 in this paper) will ensure that MHCGI works well. Therefore, the standard deviation can be set in advance easily. We also switched to a grayscale object ('Lena') to demonstrate this, as shown in Fig. 6. Obviously, a wider choice of standard deviations (0.7, 3, and 5) makes MHCGI perform better than HCGI (comparing Fig. 6(a) with 6(b)-(d)).

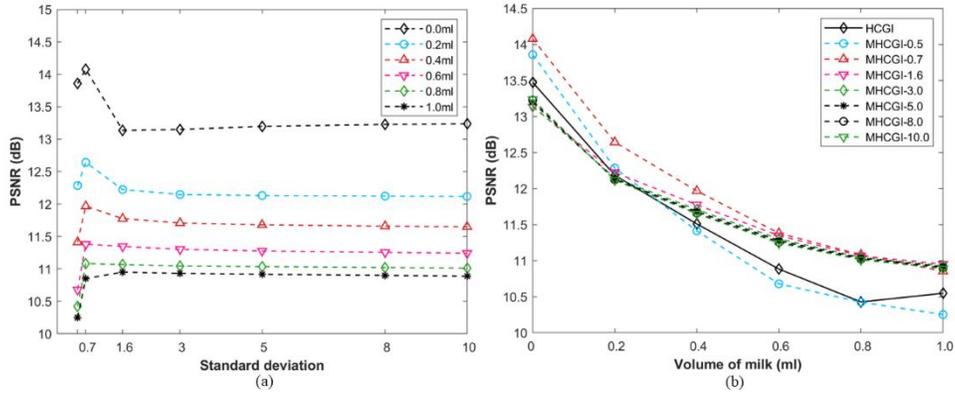

Fig. 5. Statistical curves. (a): PSNR vs. Standard deviation; (b): PSNR vs. Volume of milk; 'MHCGI-0.5' stands for the stand deviation of the 2D Gaussian filter of MHCGI is 0.5, and so on for the others.

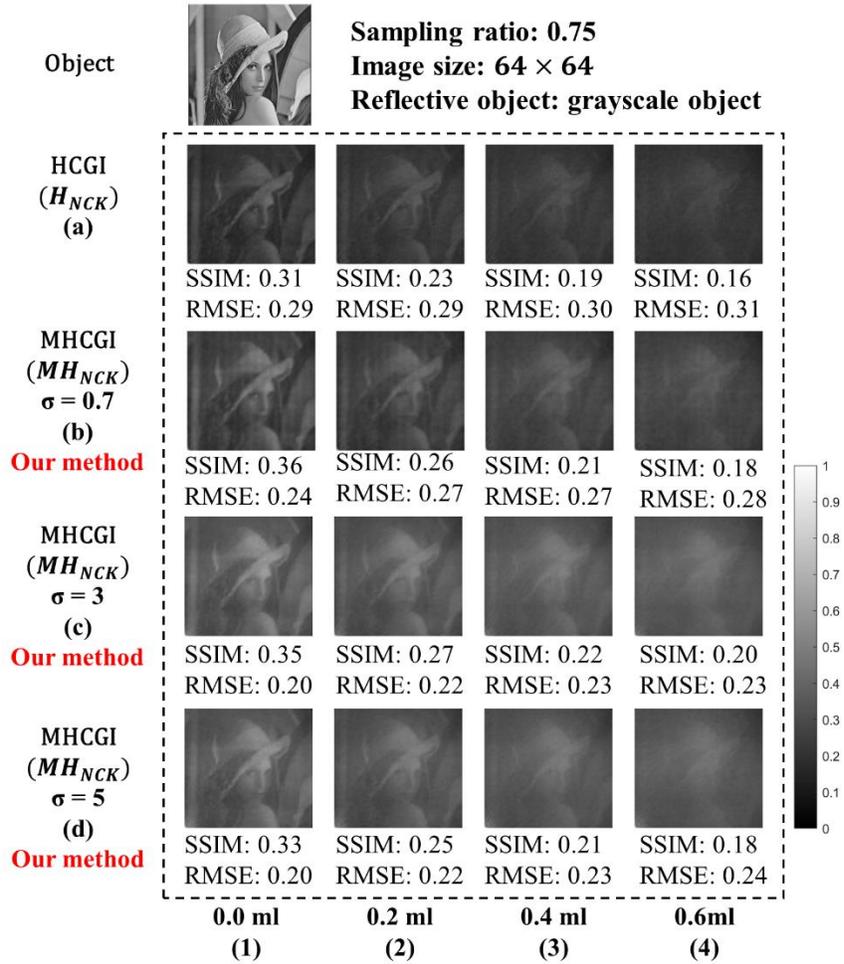

Fig. 6. Reconstruction results of the grayscale object 'Lena' at the sampling ratio of 0.75. Ambient light is kept on during the experiments. (a): the reconstruction results of HCGI; (b), (c), and (d): the reconstruction results of MHCGI with $\sigma = 0.7, 3,$ and $5$, respectively. (1)-(4): 0.0ml, 0.2ml, 0.4ml, and 0.6ml of whole milk (GB25190) are respectively added to the container in front of the object. The value of SSIM and RMSE are marked below each image.

The above experiments prove that our proposed MHCGI can enhance the robustness of CGI against turbid scattering medium. We also conducted experiments on transmissive targets (letter H) and the results also demonstrate the feasibility of our method, as shown in Fig. 7. And the details of the corresponding experiments can be seen in the third part of Supplement 1.

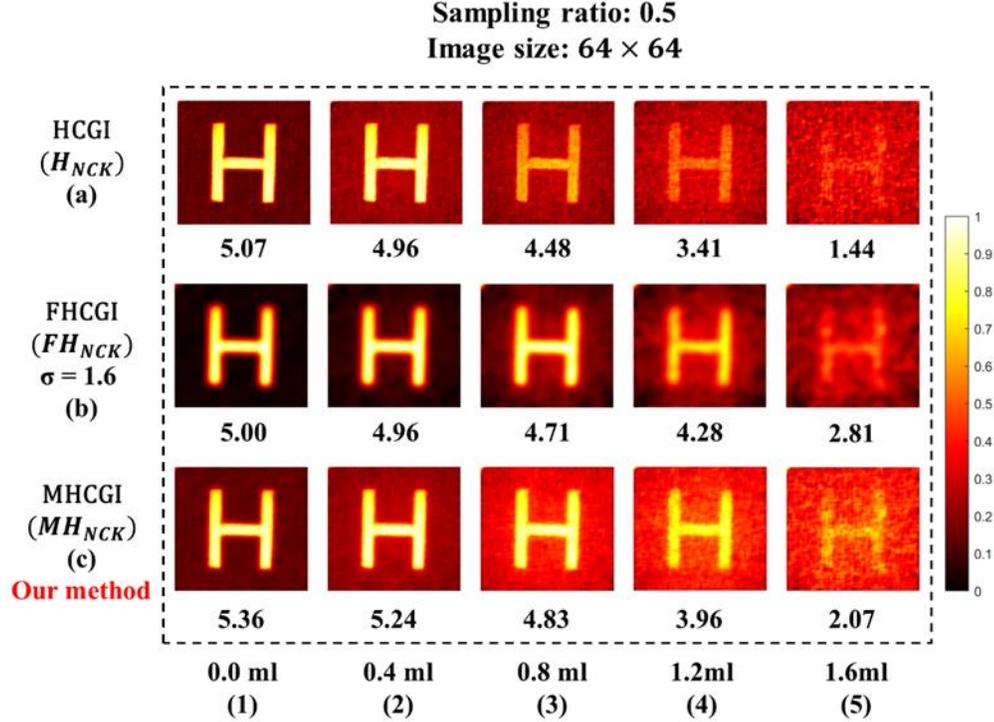

Fig. 7. Reconstruction results at the sampling ratio of 0.5. (a), (b), and (c): reconstruction results of HCGI, FHCGI, and MHCGI, respectively. The standard deviation of the 2D Gaussian filter is 1.6. The image size is 64×64 pixels. (1)-(5): 0.0ml (only pure water in the container), 0.4ml, 0.8ml, 1.2ml, and 1.6ml of whole milk (GB25190) are respectively added to the container in front of the object. The value of CNR is marked at the bottom of each image.

## 4. Discussion

Our proposed sampling matrix of $MH_{NCK}$ performs well with linear reconstruction algorithm (Eq. (15)), but does not apply to nonlinear algorithms like TVAL3 [27]. However, at a sampling ratio of 0.75, the image quality of the reconstruction results of MHCGI (linear reconstruction algorithm) is still better than the reconstruction results of the sampling matrix of $H_{NCK}$ with TVAL3, as shown in Fig. S3 in Supplement 1. (See Supplement 1 for the results). Besides, the linear reconstruction algorithm (MHCGI) consumes less reconstruction time than TVAL3.

MHCGI does not help to obtain more information about the target from the scattering environment, as it belongs to the optimization of the Hadamard sampling matrix used for image reconstruction. MHCGI directly achieves noise reduction while performing image reconstruction, unlike post-processing of images. But it is more than just noise reduction. The key is that it exploits the designable mapping relationship between patterns used for illumination and image reconstruction of CGI and combines them with the differential operation features of HCGI, as described in Step 3 of Section 2.1. Our modified Hadamard patterns achieve filtering while retaining the original high-energy components in the sampling matrix. MHCGI helps reconstruct noise-reduced results without blurred edges directly and is flexible in selecting a suitable standard deviation. It is possible to obtain high-quality reconstruction images in other scattering environments (such as fog) by combining MHCGI with other technologies that help to obtain more information about the target.

## 5. Conclusion

In this paper, we develop a kind of modified Hadamard patterns. And we analyze and demonstrate that using Hadamard patterns for illumination but using our developed modified Hadamard patterns for image reconstruction (called MHCGI in this paper) can enhance the robustness of CGI against the turbid scattering medium. The merits of our proposed MHCGI can be summarized as follows: (1) MHCGI cleverly exploits the characteristics of CGI as described in Section 4; (2) it can directly reconstruct noise-reduced results without blurred edges; (3) it is less stringent in selecting the standard deviation which allows MHCGI to set a suitable standard deviation in advance in an unknown scattering environment; (4) it is a linear reconstruction algorithm and performs better than the reconstruction results of the sampling matrix of $H_{NCK}$ with TVAL3 (nonlinear algorithm). We believe MHCGI is expected to pave the way for practical applications in imaging through more kinds of scattering mediums.

**Acknowledgments.** This work is funded by National Natural Science Foundation of China (62275022, 62275017); Beijing Municipal Natural Science Foundation (4222017).

**Supplemental document.** See Supplement 1 for supporting content.

## References

1. J. H. Shapiro, "Computational ghost imaging," Phys. Rev. A **78**, 061802(R) (2008).
2. Y. Bromberg, O. Katz, and Y. Silberberg, "Ghost imaging with a single detector," Phys. Rev. A **79**(5), 053840 (2009).
3. G. M. Gibson, S. D. Johnson, and M. J. Padgett, "Single-pixel imaging 12 years on: a review," Opt. Express **28**(19), 28190 (2020).
4. A. Valencia, G. Scarcelli, M. D'Angelo, and Y. Shih, "Two-Photon Imaging with Thermal Light," Phys. Rev. Lett. **94**(6), 063601 (2005).
5. L. Basano and P. Ottonello, "Experiment in lensless ghost imaging with thermal light," Appl. Phys. Lett. **89**(9), 091109 (2006).
6. W. Gong and S. Han, "Correlated imaging in scattering media," Opt. Lett. **36**(3), 394–396 (2011).
7. Y.-K. Xu, W.-T. Liu, E.-F. Zhang, Q. Li, H.-Y. Dai, and P.-X. Chen, "Is ghost imaging intrinsically more powerful against scattering?," Opt. Express **23**(26), 32993–33000 (2015).
8. E. Tajahuerce, V. Durán, P. Clemente, E. Irles, F. Soldevila, P. Andrés, and J. Lancis, "Image transmission through dynamic scattering media by single-pixel photodetection," Opt. Express **22**(14), 16945–16955 (2014).
9. V. Durán, F. Soldevila, E. Irles, P. Clemente, E. Tajahuerce, P. Andrés, and J. Lancis, "Compressive imaging in scattering media," Opt. Express **23**(11), 14424–14433 (2015).
10. L. Wang and S. Zhao, "Fast reconstructed and high-quality ghost imaging with fast Walsh–Hadamard transform," Photon. Res. **4**(6), 240 (2016).
11. Z. Gao, J. Yin, Y. Bai, and X. Fu, "Imaging quality improvement of ghost imaging in scattering medium based on Hadamard modulated light field," Appl. Opt. **59**(27), 8472 (2020).
12. X. Yang, Y. Liu, X. Mou, T. Hu, F. Yuan, and E. Cheng, "Imaging in turbid water based on a Hadamard single-pixel imaging system," Opt. Express **29**(8), 12010 (2021).
13. Z. Zhang, X. Wang, G. Zheng, and J. Zhong, "Hadamard single-pixel imaging versus Fourier single-pixel imaging," Opt. Express **25**(16), 19619 (2017).
14. F. Li, M. Zhao, Z. Tian, F. Willomitzer, and O. Cossairt, "Compressive ghost imaging through scattering media with deep learning," Opt. Express **28**(12), 17395 (2020).
15. J. Liu, W. Zhao, A. Zhai, and D. Wang, "Imaging through scattering media using differential intensity transmission matrices with different Hadamard orderings," Opt. Express **30**(25), 45447 (2022).
16. Z. Gao, X. Cheng, J. Yue, and Q. Hao, "Extendible ghost imaging with high reconstruction quality in strong scattering medium," Opt. Express **30**(25), 45759 (2022).
17. X.-H. Chen, F.-H. Kong, Q. Fu, S.-Y. Meng, and L.-A. Wu, "Sub-Rayleigh resolution ghost imaging by spatial low-pass filtering," Opt. Lett. **42**(24), 5290–5293 (2017).
18. K. Guo, Y. Bai, and X. Fu, "Ghost imaging of the low or high frequency based on the corresponding spatial-frequency of the reference pattern," Opt. Commun. **444**, 120–126 (2019).
19. Z. Ye, P. Zheng, W. Hou, D. Sheng, W. Jin, H.-C. Liu, and J. Xiong, "Computationally convolutional ghost imaging," Optics and Lasers in Engineering **159**, 107191 (2022).
20. B. K. Shreyamsha Kumar, "Image denoising based on gaussian/bilateral filter and its method noise thresholding," SIViP **7**(6), 1159–1172 (2013).
21. W.-K. Yu, "Super Sub-Nyquist Single-Pixel Imaging by Means of Cake-Cutting Hadamard Basis Sort," Sensors **19**(19), 4122 (2019).

22. Cai Zhuoran, Zhao Honglin, Jia Min, Wang Gang, and Shen Jingshi, "An improved Hadamard measurement matrix based on Walsh code for compressive sensing," in *2013 9th International Conference on Information, Communications & Signal Processing* (IEEE, 2013), pp. 1–4.
23. M.-J. Sun, L.-T. Meng, M. P. Edgar, M. J. Padgett, and N. Radwell, "A Russian Dolls ordering of the Hadamard basis for compressive single-pixel imaging," Sci Rep **7**(1), 3464 (2017).
24. L. López-García, W. Cruz-Santos, A. García-Arellano, P. Filio-Aguilar, J. A. Cisneros-Martínez, and R. Ramos-García, "Efficient ordering of the Hadamard basis for single pixel imaging," Opt. Express **30**(8), 13714 (2022).
25. P. G. Vaz, D. Amaral, L. F. Requicha Ferreira, M. Morgado, and J. Cardoso, "Image quality of compressive single-pixel imaging using different Hadamard orderings," Opt. Express **28**(8), 11666 (2020).
26. Z. Wang, A. C. Bovik, H. R. Sheikh, and E. P. Simoncelli, "Image Quality Assessment: From Error Visibility to Structural Similarity," IEEE Trans. on Image Process. **13**(4), 600–612 (2004).
27. C. Li, W. Yin, H. Jiang, and Y. Zhang, "An efficient augmented Lagrangian method with applications to total variation minimization," Comput Optim Appl **56**(3), 507–530 (2013).